\documentclass[usenatbib]{mnras}
\usepackage{pdflscape}
\usepackage{color}
\usepackage{IEEEtrantools} %
\usepackage{amsmath} %
\usepackage{amssymb} %
\usepackage{bm} %
\usepackage{upgreek} %
\usepackage[pdftex]{graphicx} %
\usepackage{multirow}
\usepackage{textcomp}

\def\araa{ARA\&A}
\def\apj{ApJ}
\def\apjl{ApJ}
\def\apjs{ApJS}
\def\ao{Appl.~Opt.}

\def\aap{A\&A}
\def\aapr{A\&A Rev.}

\def\mnras{MNRAS}

\def\pra{Phys.~Rev.~A}

\def\rmxaa{Revista Mexicana de Astronomia y Astrofisica}

\def\solphys{Sol.~Phys.}

\def\ssr{Space Sci.~Rev.}
\def\zap{Zeitschrift fuer Astrophysik}

\def\gca{Geochimica Cosmochimica Acta}

\newcommand{\eqn}[1]{\text{Eq.~\ref{#1}}}
\newcommand{\sect}[1]{\text{Sect.~\ref{#1}}}
\newcommand{\fig}[1]{\text{Fig.~\ref{#1}}}
\newcommand{\tab}[1]{\text{Table~\ref{#1}}}

\newcommand{\multitd}{\textsc{Multi3D}}

\newcommand{\marcs}{\textsc{MARCS}}
\newcommand{\stagger}{\textsc{Stagger}}

\newcommand{\mtd}{\textlangle3D\textrangle}

\newcommand{\kms}{\mathrm{km\,s^{-1}}}

\newcommand{\feh}{\mathrm{\left[Fe/H\right]}}

\newcommand{\xfe}[1]{\mathrm{\left[#1/Fe\right]}}
\newcommand{\lgeps}[1]{\lg{\epsilon_{\mathrm{#1}}}}
\newcommand{\sh}{S_{\mathrm{H}}}
\newcommand{\lgt}{\lg{\tau_{500}}}
\newcommand{\SiI}{\ion{Si}{I}}
\newcommand{\SiII}{\ion{Si}{II}}
\usepackage[usenames,dvipsnames,svgnames,table]{xcolor}
\newcommand{\markaschanged}[1]{#1}
\title[The solar silicon abundance]{The solar silicon abundance based on 
3D non-LTE calculations}
\author[A.~M.~Amarsi and M.~Asplund]{A.~M.~Amarsi\thanks{E-mail: 
\href{mailto:anish.amarsi@anu.edu.au}{anish.amarsi@anu.edu.au}}
and M.~Asplund\\
Research School of Astronomy and Astrophysics,
Australian National University, Canberra, ACT 2611, Australia}
\begin{document}

\date{Accepted ---. Received ---; in original form ---}
\pagerange{\pageref{firstpage}--\pageref{lastpage}} \pubyear{---}

\maketitle 
\label{firstpage}
\begin{abstract}
We present three-dimensional (3D) 
non-local thermodynamic equilibrium (non-LTE)
radiative transfer calculations for silicon in the solar photosphere,
using an extensive model atom that includes
recent, realistic neutral hydrogen collisional cross-sections. 
We find that photon losses in the \SiI~lines
give rise to slightly negative non-LTE abundance
corrections of the order $-0.01\,\mathrm{dex}$.
We infer a 3D non-LTE based solar silicon abundance
of $\lgeps{Si\odot}=7.51$.
With silicon \markaschanged{commonly chosen to be} 
the anchor between the photospheric and meteoritic abundances, 
we find that the meteoritic abundance
scale remains unchanged compared with the 
\citet{2009ARA&amp;A..47..481A}~\markaschanged{and
\citet{2009LanB...4B...44L}} results.

\end{abstract}
\section{Introduction}
\label{introduction}

Silicon is one of the most abundant metals, 
and has many astrophysical applications.
With a solar abundance and 
\markaschanged{ionisation} energy comparable to
those of iron, it is a significant electron donor in
the atmospheres of cool stars,
and a key source of opacity in the interiors of
solar-type stars. This has direct implications
on, for example, the predicted solar neutrino flux
\citep{2009ApJ...705L.123S,2016EPJA...52...78S}.
As an $\alpha$-capture element, patterns in abundance ratios such as
$\xfe{Si}$~against $\feh$, in the Milky Way
disk \citep[e.g.][]{2002A&amp;A...390..225C},
bulge \citep[e.g.][]{2016MNRAS.460..884H},
and halo \cite[e.g.][]{2007ApJ...659L.161C,
2009A&amp;A...503..533S,2013ApJ...762...26Y}, provide insight into 
stellar nucleosynthesis and the chemical evolution of the Galaxy. 
Finally, \markaschanged{silicon is commonly used
\citep[e.g.][]{2015A&amp;A...573A..25S,2015A&amp;A...573A..26S,
2015A&amp;A...573A..27G}
to set the meteoritic abundances 
\citep{2009LanB...4B...44L}~on the
same absolute scale as the solar photospheric 
abundances \citep{2009ARA&amp;A..47..481A},
because silicon is the reference element in meteorites
where hydrogen is depleted. 
(Others \citep[e.g.][]{2009LanB...4B...44L}
prefer to use a selection of elements 
to determine the scale factor
but in practice the outcome is basically
the same as when only employing 
silicon for the purpose.)}

It is therefore important to have accurate 
stellar silicon abundance determinations,
\markaschanged{and particularly for} the Sun.
\markaschanged{Unfortunately, errors can enter spectroscopic
abundance analyses from a number of different places. 
Often, errors in the transition probabilities of the 
spectral lines used to carry out the abundance analysis
have a large effect. 
This is an issue for silicon
\citep[e.g.][]{2012ApJ...755..176S},
for which few laboratory measurements have been made
within the past thirty years, 
while theoretical calculations typically have relatively large
uncertainties \citep[see for example the critical
compilation of][]{2008JPCRD..37.1285K}.
Assuming such errors are not systematic, 
which however is often the case, they can be circumvented
by basing the abundance analysis on some weighted mean 
inferred from many spectral lines.
Once given reliable transition probabilities for hopefully
many spectral abundance diagnostics,
the main systematic errors in 
the classic spectroscopic methodology arise
from the use of one-dimensional (1D) hydrostatic model atmospheres
and from the assumption that the material 
is in local thermodynamic equilibrium
\citep[LTE; e.g.][]{2005ARA&amp;A..43..481A}.}

\markaschanged{The problems with 1D hydrostatic model atmospheres stem 
from their unrealistic treatment of convection;
since they neglect fluid motions and time evolution, 1D model atmospheres
must therefore rely on the Mixing-Length Theory 
\citep[MLT;][]{1958ZA.....46..108B,1965ApJ...142..841H},
which comes with a number of free parameters that need to be calibrated.
Furthermore, spectral lines generated from 1D model atmospheres
are too narrow compared to observed line profiles
because they neglect the Doppler shifts associated with the
convective velocity field and temperature inhomogeneities,
so two more free parameters, microturbulence and macroturbulence,
must also be invoked in order to fit observed spectra
\citep[e.g][Chapter 17]{2008oasp.book.....G}.
In contrast, 3D hydrodynamical model solar and stellar atmospheres
successfully reproduce the observations to exquisite detail,
including the line shapes, shifts and asymmetries
\citep[e.g.][]{2000A&amp;A...359..729A,
2009LRSP....6....2N,2013A&amp;A...554A.118P}.}

There have been several detailed investigations into the departures
from LTE in \SiI~lines in the solar photosphere.
Non-LTE \markaschanged{calculations} based on 1D model atmospheres
by \citet{2001A&amp;A...373..998W}~found
non-LTE abundance corrections ($\lgeps{NLTE}-\lgeps{LTE}$)~that 
are typically very slightly negative, 
and of the order $-0.01\,\mathrm{dex}$,
a result later consolidated by \citet{2008A&amp;A...486..303S}~using
\markaschanged{a more extensive} model atom.
The non-LTE calculations by \citet{2012KPCB...28..169S}~using 
1D hydrostatic model atmospheres, and 
by \cite{2012ApJ...755..176S}~using a 3D hydrodynamic model atmosphere
and treating each column of the model atmosphere
independently (i.e.~the so-called 1.5D~approximation),
suggest slightly more severe abundance corrections
of the order $-0.05\,\mathrm{dex}$.
Beyond the Sun, there is evidence of much larger
non-LTE effects in \SiI~lines~\citep[e.g.][]{2009A&amp;A...503..533S,
2011A&amp;A...534A.103S,
2012ApJ...755...36S,2013ApJ...764..115B,2016ApJ...823...36T}.

Recent 1D non-LTE calculations by \citet{2016AstL...42..366M}~suggest
that the severity of the non-LTE effects in the afore-mentioned studies
may have been overestimated. 
\citet{2016AstL...42..366M}~utilized for the first time
the collisional cross-sections 
of \citet{2014A&amp;A...572A.103B}~for
excitation and charge-transfer with neutral hydrogen,
which were calculated using the Born-Oppenheimer formalism.
Using the model atom of \citet{2008A&amp;A...486..303S},
and 1D \marcs~hydrostatic model atmospheres \citep{2008A&amp;A...486..951G},
\citet{2016AstL...42..366M}~commented briefly that
the new collisional data reduce the non-LTE effects
to vanishingly small levels in metal-poor turn-off stars.
This is a significant result because,
prior to these cross-sections becoming available,
the semi-empirical recipe 
of \citet{1968ZPhy..211..404D,1969ZPhy..225..483D},
as formulated by
\citet[][]{1984A&amp;A...130..319S}~or
\citet[][]{1993PhST...47..186L},
was used, typically with a global scaling factor $\sh=0.1$.
This recipe does not provide a \markaschanged{realistic}
description of the physics of the collisional interactions,
being based on
the classical \cite{thomson1912xlii}~electron 
\markaschanged{ionisation} cross-section;
it is typically in error by several orders of magnitude
\citep[e.g.][]{2016A&amp;ARv..24....9B}.

The canonical solar photospheric silicon abundance itself has 
seen a slight downwards revision, from
$\lgeps{Si\odot}=7.55$~\citep{1989GeCoA..53..197A,1998SSRv...85..161G},
to
$\lgeps{Si\odot}=7.51$~\citep{2000A&amp;A...359..755A,
2005ASPC..336...25A,2009ARA&amp;A..47..481A,
2015A&amp;A...573A..25S}.
The most recent of these was based on a 3D LTE analysis
of nine \SiI~lines and one \SiII~line 
\markaschanged{in the solar disk-centre intensity spectrum}
and adopting
1D non-LTE abundance corrections 
\markaschanged{to the solar flux spectrum} 
from \citet{2008A&amp;A...486..303S}.
Correcting 3D LTE abundances with 1D non-LTE abundance
corrections \markaschanged{in this way} is not consistent; 
furthermore these abundance corrections were obtained
prior to the calculations of \citet{2014A&amp;A...572A.103B}~\markaschanged{for
the neutral hydrogen collisional cross-sections.}
A detailed investigation
using a 3D hydrodynamic model solar atmospheres
and 3D non-LTE radiative transfer, with the best available
atomic data, is therefore highly desirable. 

In this paper we study the 3D non-LTE \SiI~line formation
in a 3D hydrodynamic model solar atmosphere.
To obtain accurate results we construct a realistic
model atom that includes recent neutral hydrogen collision data
from \citet{2014A&amp;A...572A.103B}. 
We use the same \stagger~model solar atmosphere 
as that used by \citet{2015A&amp;A...573A..25S}.
This enables us to directly apply our derived abundance corrections 
to their 3D LTE results, and thereby obtain
a consistent 3D non-LTE solar photospheric silicon abundance.


\section{Method}
\label{method}

\subsection{3D non-LTE radiative transfer code}
\label{methodcode}
\begin{figure}
\begin{center}
\includegraphics[scale=0.3]{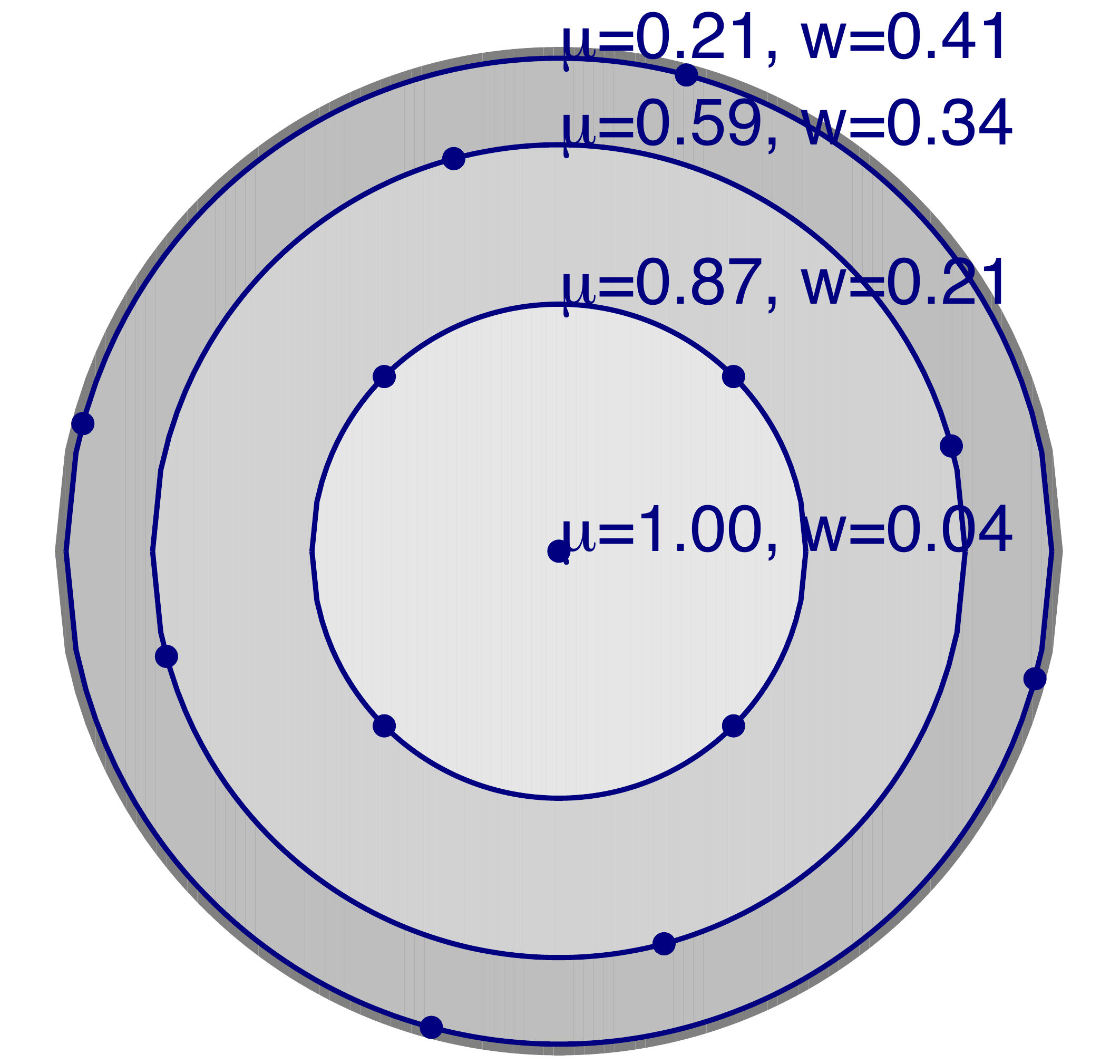}
\caption{Projection of the unit sphere onto a 2D plane,
illustrating the angle quadrature adopted for the
statistical equilibrium calculations. 
The rings~of constant $\mu$~and their 
weights $w$~are set by the $8\text{-point}$~Lobatto quadrature.
The~$\phi$~nodes on a given ring of constant $\mu$~are
distributed evenly, and are given
equal weight (i.e.~equidistant trapezoidal quadrature).
The $\phi$~nodes on different rings of constant $\mu$~are 
rotated relative to each other by $\frac{\uppi}{6}$~radians.}
\label{anglequadrature}
\end{center}
\end{figure}

A customized version of \multitd~\citep{2009ASPC..415...87L}~was
used to solve for the statistical equilibrium (and to 
subsequently compute the emergent spectra), 
under the assumption that silicon is a
trace element with no influence on the background 
pseudo-static model atmosphere
\citep[the so-called restricted non-LTE problem;][]{1971ARA&amp;A...9..237H}.
We refer the reader to the methodologies outlined in
\citet{paperiii,2016MNRAS.455.3735A}~for
further details about the code.

We illustrate the angle quadrature
that was used to calculate the mean radiation field
during the statistical equilibrium calculations,
in \fig{anglequadrature}.
A finer angle quadrature was adopted in this study
than in the two afore-mentioned papers.
For the integral over $\mu=\cos\theta$,
an $8\text{-point}$~Lobatto quadrature on the interval [$-1,1$]~was adopted,
and for the integral over $\phi$~for the non-vertical rays, 
an equidistant $4\text{-point}$~trapezoidal integration 
on the interval [$0,2\uppi$]~was adopted.
This equates to $13\,\text{outgoing rays}$~on
the unit hemisphere (\fig{anglequadrature}),
or $26\,\text{rays}$~over the unit sphere in total.

\subsection{Model atmospheres}
\label{methodatmosphere}
The model solar atmosphere used in this study was 
first presented by \citet{2009ARA&amp;A..47..481A},
and later tested against observational constraints by
\citet{2013A&amp;A...554A.118P}; it is the same model atmosphere
used in the ongoing solar chemical composition series
\citep{2015A&amp;A...573A..25S,2015A&amp;A...573A..26S,
2015A&amp;A...573A..27G}.
It was computed using 
the \stagger~code~\citep{nordlund19953d,1998ApJ...499..914S},
albeit with some customizations
\citep[e.g.][]{2011JPhCS.328a2003C,2013A&amp;A...557A..26M}.
We refer the reader to those studies for further details 
about the hydrodynamical simulations.

The original Cartesian mesh of $240\times240\times230$~(excluding
the five ghost zones on the top and bottom of the simulation box)
was reduced to $120\times120\times101$~for the
3D non-LTE calculations, to save computing time.
This was done by selecting every second \markaschanged{grid point in 
each of the horizontal dimensions},
trimming optically deep layers (i.e.~removing layers with
vertical optical depth $500\,\mathrm{nm}$: $\lgt\gtrsim3.0$).
\markaschanged{To test the impact of degrading the horizontal resolution,
a 3D non-LTE test calculation was carried out using a single snapshot
with $60\times60\times101$~resolution, and the equivalent widths were 
compared with those obtained from the same snapshot with
$120\times120\times101$~resolution. 
The differences in the equivalent widths
in the vertical intensity were minimal
(of the order $0.00001\,\mathrm{dex}$~in the worst case).
The differences in the equivalent widths
in the inclined intensities were slightly larger
(of the order $0.003\,\mathrm{dex}$~in the worst case),
but still negligible for our purposes.
The analysis presented in this work is based on abundance corrections
derived from the equivalent widths in the vertical intensity.
Hence we conclude that downgrading the horizontal resolution by 
a factor of four from $240^{2}$~to $120^{2}$~has 
no influence on the conclusions presented in this work.}

Calculations were performed on six snapshots of the
numerically-relaxed section of the full simulation,
equidistant across $45.5\,\mathrm{minutes}$ of solar time.
\markaschanged{This number of snapshots was sufficient to obtain
3D non-LTE versus 3D LTE abundance corrections to a precision
of better than $0.001\,\mathrm{dex}$, which could then be applied
to the 3D LTE abundance derived by \citet{2015A&amp;A...573A..25S}~that
was based on a full sequence of snapshots.
This time-sampling error of less than $0.001\,\mathrm{dex}$~in
the abundance corrections was determined by comparing 
the abundance corrections derived from the individual snapshots 
with the final value obtained after integrating over 
the selected snapshots.}
The 3D non-LTE calculations were performed for three silicon abundances:
$\lgeps{Si\odot}=\{7.39\text{, }7.51\text{, }7.63\}$;
the background chemical composition remained fixed using
the abundances presented in \citet{2009ARA&amp;A..47..481A}.
\markaschanged{This resolution in silicon abundance was also sufficient
to obtain interpolated 3D non-LTE versus 3D LTE abundance corrections to
a precision of better than $0.001\,\mathrm{dex}$.}

A horizontally- and temporally-averaged 3D model solar atmosphere
(henceforth \mtd) was constructed by
averaging the gas temperature and
logarithmic gas density from the 3D model atmosphere
on surfaces of equal time and vertical optical depth at $500\,\mathrm{nm}$.
All other quantities were then calculated
consistently via the equation of state.
In the \mtd~model atmosphere the velocities
were set to zero everywhere.
As such, a microturbulent broadening parameter 
needed to be included in the line-formation calculations
to account for the line broadening caused by convective motions;
$\xi=1.0\,\kms$~was adopted.
We emphasize that these broadening effects are
naturally taken into account when performing
line-formation calculations in the 3D model atmosphere,
without having to envoke any microturbulent broadening parameters
\citep[e.g.][]{2000A&amp;A...359..729A}.

\subsection{Model atom}
\label{methodatom}
\begin{figure*}
\begin{center}
\includegraphics[scale=0.65]{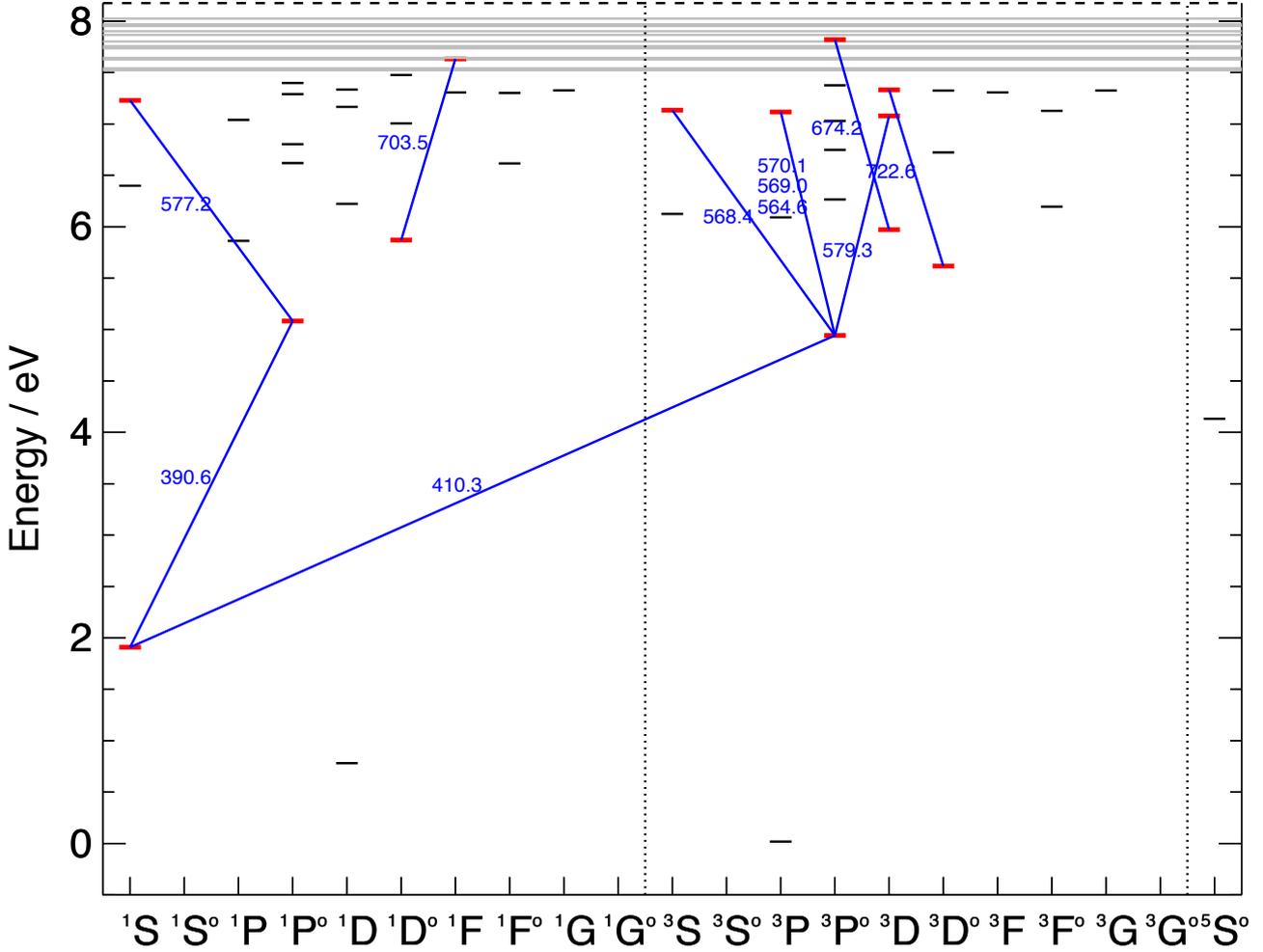}
\caption{Grotrian diagram of the 
reduced model atom for which the statistical equilibrium
was solved. 
Levels are shown in black;
levels corresponding to the lines
used in the subsequent analysis are shown in
red, and these lines are also shown
and labelled using their wavelengths in air in nanometres
(c.f.~\tab{abundancelines}).
Twelve super levels are indicated in grey at the top of the diagram.}
\label{grotrian}
\end{center}
\end{figure*}

\begin{table}
\begin{center}
\caption{\SiI~lines analysed in this work. 
The nine intermediate-excitation 
optical lines, and their parameters, correspond to those
used by by \citet{2015A&amp;A...573A..25S}~in
their solar analysis.
Two low-excitation \markaschanged{violet} \SiI~lines were also 
included in this study, but were not considered
in the final solar abundance analysis.
The oscillator strengths are those measured by \citet{1973A&amp;A....26..471G},
increased by $0.097\,\mathrm{dex}$~after
renormalisation using the accurate lifetimes
measured by \citet{1991PhLA..152..407O,1991PhRvA..44.7134O}.}
\label{abundancelines}
\begin{tabular}{c r l c c}
\hline
$\lambda_{\text{Air}} / \mathrm{nm}$ &
\multicolumn{1}{c}{Lower level} & 
\multicolumn{1}{c}{Upper level} &
$E_{\text{low}}$ & 
$\log{g\,f}$ \\
\hline
\hline
$390.553$ & 
$3s^{2}\,3p^{2} \,\, {^{1}\mathrm{S}^{\text{}}_{0}}$ &
$3s^{2}\,3p\,4s \,\, {^{1}\mathrm{P}^{\text{o}}_{1}}$ &
$1.9087$ & 
$-0.99$ \\

$410.293$ & 
$3s^{2}\,3p^{2} \,\, {^{1}\mathrm{S}^{\text{}}_{0}}$ &
$3s^{2}\,3p\,4s \,\, {^{3}\mathrm{P}^{\text{o}}_{1}}$ &
$1.9087$ & 
$-3.04$ \\

\hline
$564.561$ & 
$3s^{2}\,3p\,4s \,\, {^{3}\mathrm{P}^{\text{o}}_{1}}$ &
$3s^{2}\,3p\,5p \,\, {^{3}\mathrm{P}^{\text{}}_{2}}$ &
$4.9296$ & 
$-2.04$ \\

$568.448$ & 
$3s^{2}\,3p\,4s \,\, {^{3}\mathrm{P}^{\text{o}}_{2}}$ &
$3s^{2}\,3p\,5p \,\, {^{3}\mathrm{S}^{\text{}}_{1}}$ &
$4.9538$ & 
$-1.55$ \\

$569.043$ & 
$3s^{2}\,3p\,4s \,\, {^{3}\mathrm{P}^{\text{o}}_{1}}$ &
$3s^{2}\,3p\,5p \,\, {^{3}\mathrm{P}^{\text{}}_{1}}$ &
$4.9296$ & 
$-1.77$ \\

$570.111$ & 
$3s^{2}\,3p\,4s \,\, {^{3}\mathrm{P}^{\text{o}}_{1}}$ &
$3s^{2}\,3p\,5p \,\, {^{3}\mathrm{P}^{\text{}}_{0}}$ &
$4.9296$ & 
$-1.95$ \\

$577.215$ & 
$3s^{2}\,3p\,4s \,\, {^{1}\mathrm{P}^{\text{o}}_{1}}$ &
$3s^{2}\,3p\,5p \,\, {^{1}\mathrm{S}^{\text{}}_{0}}$ &
$5.0823$ & 
$-1.65$ \\

$579.307$ & 
$3s^{2}\,3p\,4s \,\, {^{3}\mathrm{P}^{\text{o}}_{1}}$ &
$3s^{2}\,3p\,5p \,\, {^{1}\mathrm{D}^{\text{}}_{2}}$ &
$4.9296$ & 
$-1.96$ \\

$674.164$ & 
$3s^{2}\,3p\,4p \,\, {^{3}\mathrm{D}^{\text{}}_{3}}$ &
$3s^{2}\,3p\,8s \,\, {^{3}\mathrm{P}^{\text{o}}_{2}}$ &
$5.9840$ & 
$-1.65$ \\

$703.490$ & 
$3s^{2}\,3p\,3d \,\, {^{1}\mathrm{D}^{\text{o}}_{2}}$ &
$3s^{2}\,3p\,5f \,\, {^{1}\mathrm{F}^{\text{}}_{3}}$ &
$5.8708$ & 
$-0.78$ \\

$722.621$ & 
$3s\,3p^{3} \,\, {^{3}\mathrm{D}^{\text{o}}_{1}}$ &
$3s^{2}\,3p\,4f \,\, {^{3}\mathrm{D}^{\text{}}_{2}}$ &
$5.6135$ & 
$-1.41$ \\
\hline
\hline
\end{tabular}
\end{center}
\end{table}

\subsubsection{Overview}
\label{methodatomoverview}

The atomic physics of the non-LTE species is encapsulated
in the model atom.
The model atom needs to be complete and realistic 
in order to obtain reliable abundance corrections
and an accurate solar photospheric silicon abundance.
At the same time, the model atom needs to be small enough
to permit 3D non-LTE calculations on the large 
model solar atmosphere used in this study.
Similar to \citet{2008ApJ...682.1376B},
our approach was to first construct a comprehensive model atom
using the best available data (\sect{methodatomlarge}),
and to subsequently reduce its complexity
by collapsing fine structure levels and merging 
high-excitation levels into super levels (\sect{methodatomsmall}),
while ensuring that the results from the reduced model atom
remained consistent with those from the comprehensive model atom;
these \markaschanged{tests} were done using the 
\mtd~solar model atmosphere.

We illustrate the final, reduced model atom in \fig{grotrian}.
It consists of $56\,\text{levels}$~of \SiI~plus
the ground state of 
\SiII~and $634\,\text{radiative bound-bound transitions}$;
all 
\markaschanged{photoionisations}
between \SiI~levels and the ground state of \SiII~are
included. As with the comprehensive model atom 
used by \citet{2008ApJ...682.1376B},
the model atom does not include any \SiII~levels
above the ground state. 
The first excited state of \SiII~is $5.3\,\mathrm{eV}$~above the 
ground state, which means that their collisional coupling
with \SiI~levels is minute. 
The excited states are also sparsely populated in the solar photosphere
\citep[e.g.][]{2001A&amp;A...373..998W},
implying that any non-LTE behaviour in the \SiII~ion
will have only a minute effect on the ground state of \SiII.
Consequently, the excited states of \SiII~are not expected 
to have any significant influence on the populations of \SiI.

After the statistical equilibrium was solved for the reduced model atom,
the populations were redistributed onto another model atom
that had fine structure resolved, as described in \citet{paperiii}.
The emergent spectra were calculated for the 
(fine structure)~lines listed in \tab{abundancelines},
which correspond to the nine intermediate-excitation 
\SiI~lines~used by \citet{2015A&amp;A...573A..25S}~in
their solar analysis, as well as two
low-excitation \markaschanged{violet} \SiI~lines 
\markaschanged{that can be used the 
as the diagnostics for the structure of the solar chromosphere
\citep[][]{2001ApJ...552..877C}~and are also}
commonly used \markaschanged{to determine
the abundance of silicon in} metal-poor stars
\citep[e.g.][]{2007ApJ...659L.161C,2013ApJ...762...26Y}.

\markaschanged{The oscillator strengths of the 
lines listed in \tab{abundancelines} are based
on those measured by \citet{1973A&amp;A....26..471G}
and renormalised using the accurate lifetimes
measured by \citet{1991PhLA..152..407O,1991PhRvA..44.7134O}.
A complication arises because they do not use LS coupling for the
upper levels of three of the lines used here: the $674.2\,\mathrm{nm}$, 
$703.5\,\mathrm{nm}$, and $722.6\,\mathrm{nm}$~lines.
For the first two of these levels, the corresponding LS coupling level
can be uniquely determined via the LS coupling selection rules.
For the third level, there is a choice of 
$^{3}D_{2}$~and $^{3}F_{2}$~levels;
the former was adopted. 
The difference in the departure coefficients between these
two levels in the line forming regions 
$-2\lesssim\lgt\lesssim0$~is less than $0.01\%$~in 
the 3D model atmosphere snapshots;
this implies that the choice has a negligible impact on the results.}

\subsubsection{Comprehensive model atom}
\label{methodatomlarge}

Following \citet{2008ApJ...682.1376B}, the main source of
energies, oscillator strengths, and 
\markaschanged{photoionisation} cross-sections
was the Opacity Project online database 
\citep[TOPBASE;][]{1992RMxAA..23..107C,1993A&amp;A...275L...5C}.
This data was computed under the assumption
of strict LS coupling, and without any resolution of
fine structure. The agreement with
observed energies is typically at the 1\%~level,
while the uncertainties in the oscillator strengths
and \markaschanged{photoionisation} cross-sections
are typically on the 10\%~level
\citep{1993JPhB...26.1109N}.

The TOPBASE data set is relatively complete; a few missing Rydberg states
with electron configurations of the form $3s^{2}\,3p\,nl$~were
computed using the Rydberg formula, 
\phantomsection\begin{IEEEeqnarray}{rCl}
    E-E_{\infty}&=&-\frac{\mathrm{Ry}}{\left(n-\delta_{l}\right)^{2}}\, ,
\end{IEEEeqnarray}
up to $0.2\,\mathrm{eV}$~below the 
\markaschanged{ionisation} limit,
using a fit to the data from TOPBASE to roughly estimate the
quantum defects $\delta_{l}$.
Tests revealed that the presence of these extrapolated 
levels \markaschanged{does} not have a significant effect on the main 
findings of this paper;
nevertheless they were retained in the model atom.
Missing 
\markaschanged{photoionisation} cross-sections
(including those for extrapolated levels) were estimated
using a hydrogenic expression with Gaunt factors
from \citet{1935MNRAS..96...77M}~as 
given in \citet[][Chapter 8]{2008oasp.book.....G}.

The data was refined where possible using 
observed fine-structure energies from \citet{1983JPCRD..12..323M}
and fine-structure oscillator strengths from various sources
via the NIST online database \citep{NIST_ASD}.
As with the lines listed in \tab{abundancelines},
oscillator strengths from \citet{1973A&amp;A....26..471G},
were increased by $0.097\,\mathrm{dex}$~after
renormalisation using the accurate lifetimes
measured by \citet{1991PhLA..152..407O,1991PhRvA..44.7134O}.

The rate coefficients for excitation via electron collisions
were calculated using the semi-empirical recipe of
\citet{1962ApJ...136..906V}.
The Einstein coefficient for spontaneous emission
enters into this recipe; 
for radiatively forbidden transitions
the Einstein coefficient was calculated
by assuming an effective oscillator strength $f_{\text{forb}}=0.001$. 
For \markaschanged{ionisation} via electron collisions
the empirical formula given in 
\citet[][Chapter 3]{1973asqu.book.....A}~was adopted.

The rate coefficients for excitation and 
charge transfer via neutral hydrogen collisions
involving low- and intermediate-excitation \SiI~levels
were taken from the calculations
presented by \citet{2014A&amp;A...572A.103B},
based on the Born-Oppenheimer formalism.
For charge transfer from high-excitation levels,
and for excitation from low-excitation levels
to high-excitation levels,
the rate coefficients were obtained via fits in the 
$\lg{\left(\text{rate}\right)}\,
\text{versus}\,\lg{\left(\Delta\,E\right)}$~plane.
The excitation rate coefficients 
involving high-excitation levels were calculated
via the free-electron model of 
\citet{1985JPhB...18L.167K,
kaulakys1986free,1991JPhB...24L.127K}~in
the scattering length approximation,
using the routines presented by \citet{2016A&amp;ARv..24....9B}.

Finally, collisional transitions within the same \markaschanged{term}
were set to extremely large values \markaschanged{to ensure that the 
corresponding fine-structure levels are
populated according to their statistical weights and have 
identical departure coefficients}
\citep[e.g.][]{1993A&amp;A...275..269K}.

\subsubsection{Reduced model atom}
\label{methodatomsmall}

Since the comprehensive model atom
consists of over $50000$~frequency points,
\markaschanged{it} was necessary to reduce the size of the model atom
in order to make the 3D non-LTE calculations computationally
tractable. To proceed, 
fine structure levels were collapsed into single levels.
Following \citet{martin1999atomic}\footnote{\url{http://www.nist.gov/pml/pubs/atspec/index.cfm}},
the average statistical weight and 
energy of a collapsed \markaschanged{term}~$I$~are given by,
\phantomsection\begin{IEEEeqnarray}{rCl}
    \label{eq_fsg}
    \overline{g}_{I}&=&\displaystyle\sum\limits_{i\in I} g_{i}\, ,\\
    \label{eq_fse}
    \overline{E}_{I}&=&\displaystyle\sum\limits_{i\in I} 
    \frac{g_{i}\,E_{i}}{\overline{g}_{I}}\, ,
\end{IEEEeqnarray}
and the average wavelength and oscillator strength 
\markaschanged{of a multiplet} between two collapsed
\markaschanged{terms} $I$, $J$~are given by, 
\phantomsection\begin{IEEEeqnarray}{rCl}
    \label{eq_fsl}
    \overline{\lambda}_{I\,J}&=&\frac{h\,c}
    {\overline{E}_{J}-\overline{E}_{I}}\, ,\\
    \label{eq_fsf}
    \overline{f}_{I\,J}&=&
    \displaystyle\sum\limits_{i\in I,j\in J} 
    \frac{g_{i}\,\lambda_{i\,j}\,f_{i\,j}}
    {\overline{g}_{I}\,\overline{\lambda}_{I\,J}}\, .
\end{IEEEeqnarray}

Under the assumption that closely separated levels
are collisionally coupled and thus have identical departure coefficients,
levels above $7.5\,\mathrm{eV}$~separated by 
up to $0.04\,\mathrm{eV}$~were
merged into super-levels, and affected radiative transitions
were merged into super-transitions
\citep[e.g.][Chapter 18]{2014tsa..book.....H}.
This merging was analogous to the collapsing of fine structure
(\eqn{eq_fsg}~to \eqn{eq_fsf}),
albeit with levels weighted by their
Boltzmann factors 
$g_{i}\exp{\left(-E_{i}/k_{\mathrm{B}}\,T\right)}$~and
using a characteristic temperature of $T=5000\,\mathrm{K}$,
rather than simply their statistical weights $g_{i}$.
Levels corresponding to lines to be used for the abundance
analysis were not merged into super-levels
(as illustrated in \fig{grotrian}).

Lines with wavelengths greater than $10\,\mathrm{\mu m}$ 
were cut from the model atom.
These lines do not significantly alter the statistical equilibrium, 
because they correspond
to levels that are separated in energy
by less than about $0.124\,\mathrm{eV}$;
such levels are in close collisional coupling in the solar photosphere.

Tests on the \mtd~model solar atmosphere
revealed the error in the non-LTE 
\markaschanged{equivalent widths in the vertical intensity}
incurred by collapsing the comprehensive model atom
is less than $0.001\,\mathrm{dex}$.
We suspect that the error incurred in the final 3D non-LTE results
from collapsing the atom
is insignificant compared to other uncertainties inherent in the 
model atom and hydrodynamical model atmosphere,
and in the non-LTE radiative transfer calculations themselves.

\section{Line formation in the solar photosphere}
\label{results}

\subsection{Non-LTE effect}
\label{resultsnonlte}
\begin{figure*}
\begin{center}
\includegraphics[scale=0.65]{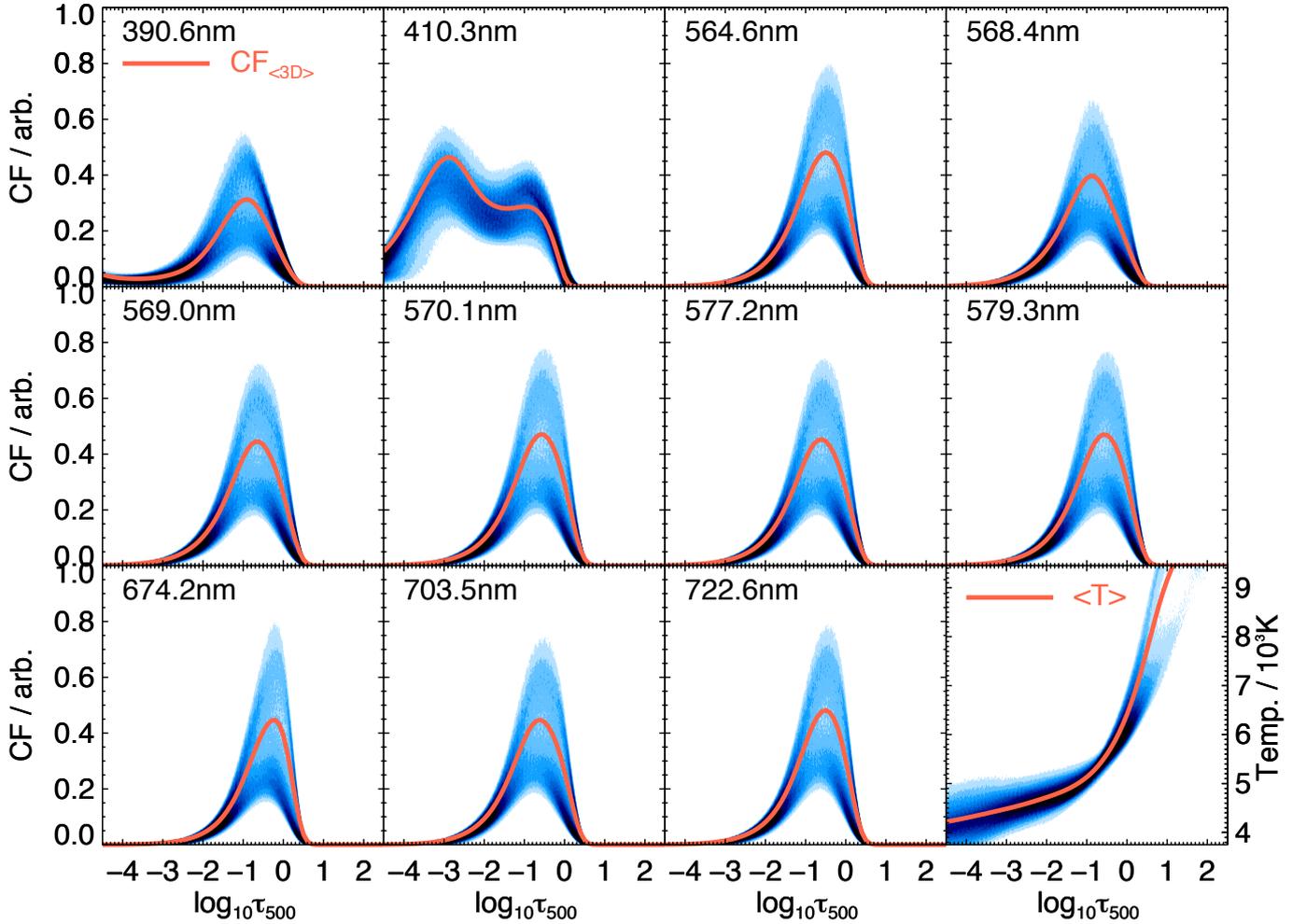}
\caption{Non-LTE contribution functions
to the line depression 
\markaschanged{in the vertical intensity,
integrated over frequency $\nu$:
$\int{\alpha_{\nu}\,S^{\text{eff}}_{\nu}\,\mathrm{e}^{-\tau_{\nu}}\,
\mathrm{d}\nu}$, 
where the vertical optical depth is $\tau_{\nu}$,
the effective source function is
$S^{\text{eff}}_{\nu}=
\left(\alpha^{\mathrm{l}}_{\nu}/\alpha_{\nu}\right)\left(I^{\mathrm{c}}_{\nu}-
S^{\mathrm{l}}_{\nu}\right)$,
$\alpha_{\nu}$~and $S_{\nu}$~are the 
linear extinction coefficient and source function and
$\mathrm{l}$~and $\mathrm{c}$~denote
line and continuum quantities, respectively.
This is shown for}
the eleven lines listed in \tab{abundancelines},
in arbitrary units (contours).
Also shown \markaschanged{are the corresponding contribution functions
in the \mtd~model atmosphere: $\mathrm{CF}_{\mathrm{\langle 3D\rangle}}$}. 
For perspective, the bottom right panel
shows the temperature stratification
of the 3D and \mtd~model solar atmospheres.}
\label{contributionfunction}
\end{center}
\end{figure*}

\begin{figure*}
\begin{center}
\includegraphics[scale=0.65]{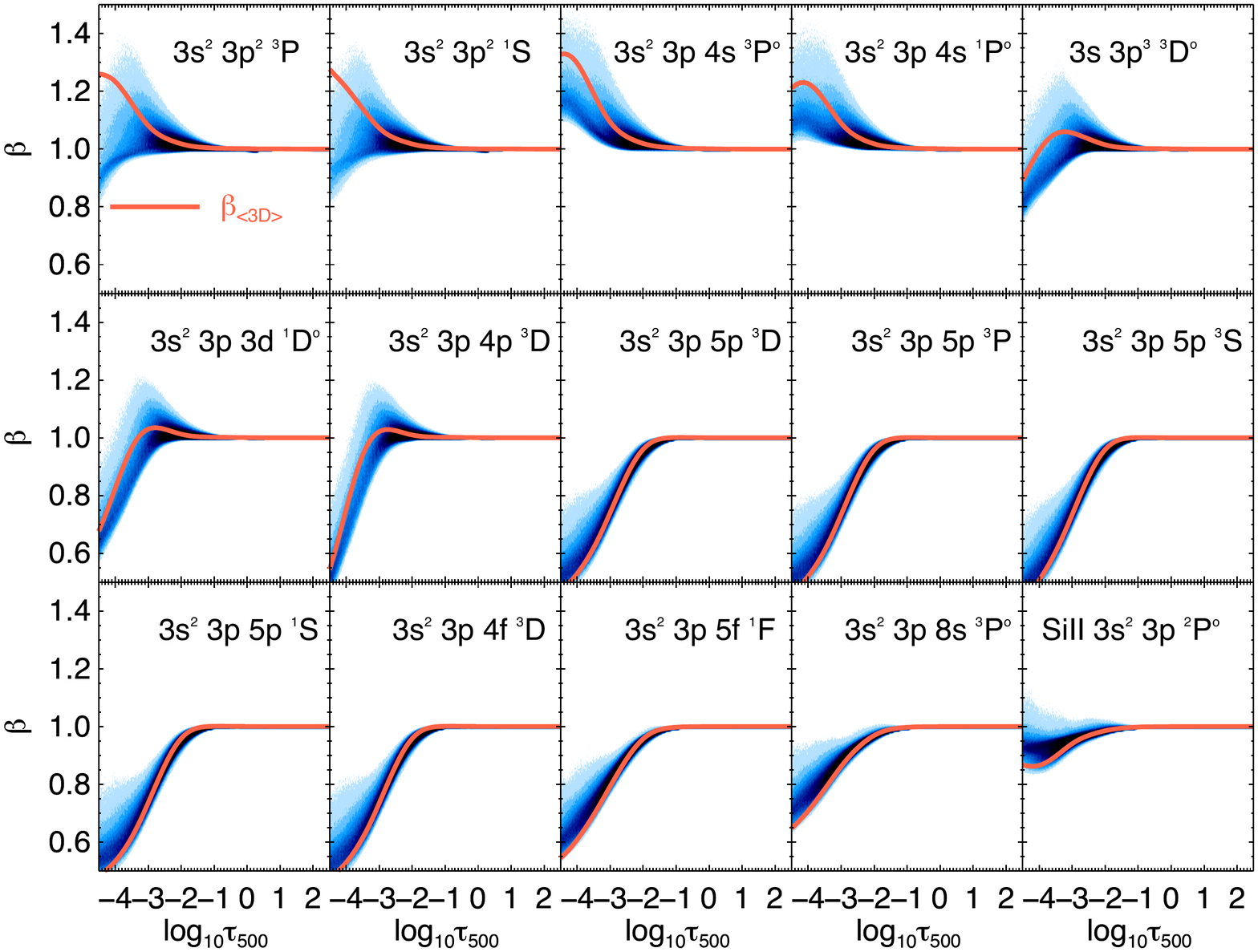}
\caption{Departure coefficients
$\beta=n^{\mathrm{NLTE}}/n^{\mathrm{LTE}}$~for the
thirteen levels listed in \tab{abundancelines},
as well as the ground states of 
\SiI~and \SiII,
in the 3D model atmosphere (filled contours).
Also shown \markaschanged{are the 
corresponding departure coefficients
in the \mtd~model atmosphere:
$\beta_{\mathrm{\langle 3D\rangle}}$}.}
\label{departurecoefficient}
\end{center}
\end{figure*}

\begin{figure*}
\begin{center}
\includegraphics[scale=0.3]{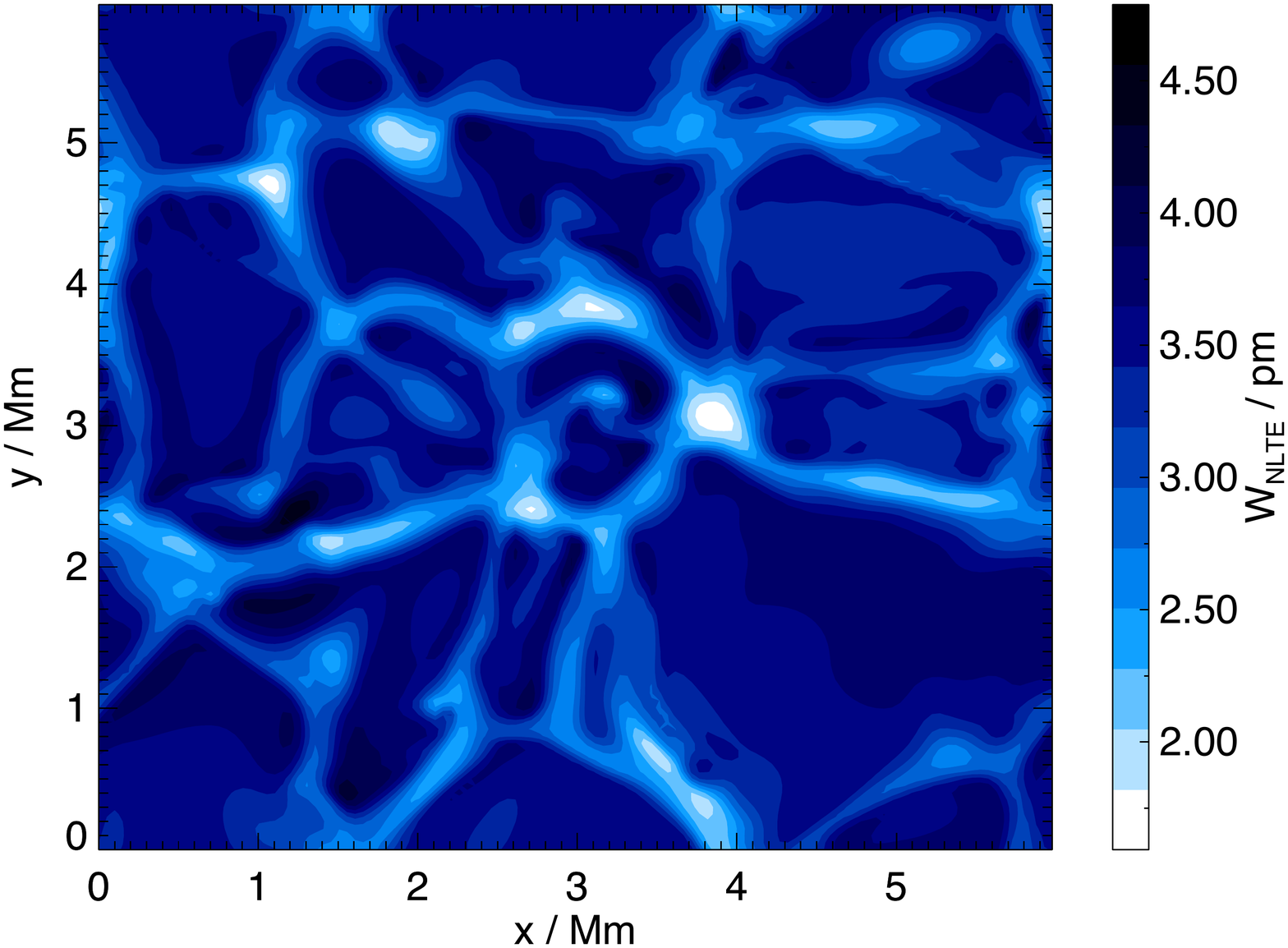}
\includegraphics[scale=0.3]{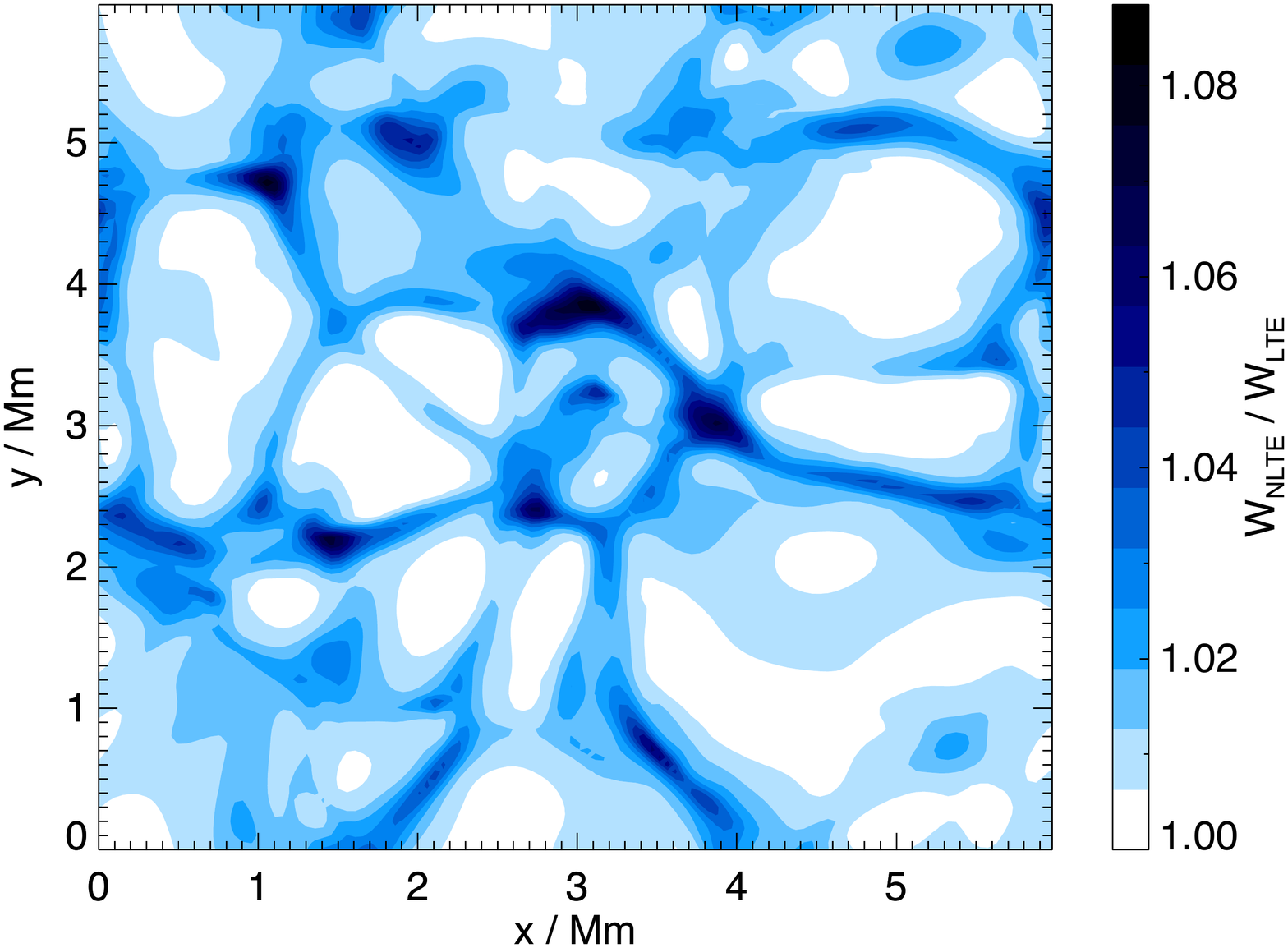}
\caption{Non-LTE equivalent widths 
\markaschanged{in the vertical intensity} (left),
and non-LTE to LTE equivalent width ratios 
\markaschanged{in the vertical intensity} (right),
for the $564.6\,\mathrm{nm}$~line
on the surface of a snapshot of the
3D~model solar atmosphere for
a fixed silicon abundance $\lgeps{Si}=7.51$.
The ratios are always greater than one or equal to one, indicating
that the emergent equivalent widths are systematically 
stronger in non-LTE.
The equivalent widths and equivalent width ratios 
for the other \SiI~lines listed in \tab{abundancelines}~take on
a similar appearance.}
\label{map}
\end{center}
\end{figure*}

To understand the non-LTE effect in individual lines,
the line-forming regions must first be identified. 
We illustrate the \SiI~line-forming regions 
\markaschanged{in the vertical intensity}
in \fig{contributionfunction}. The intermediate-excitation \SiI~lines
mostly form between $-2\lesssim\lgt\lesssim0$.
The two low-excitation \SiI~lines are saturated
with significant optical depths even at the
very top layers of the model atmosphere. 
This indicates that the solar model atmosphere may not be extended enough
for these two lines, and thus that
their absolute equivalent widths may not be reliable
(although qualitative results for these lines remain useful);
they were not considered in the 
abundance analysis of \cite{2015A&amp;A...573A..25S},
and are consequently not considered in the 
abundance analysis presented here either.

We illustrate the departure coefficients,
$\beta=n_{\text{NLTE}}/n_{\text{LTE}}$, in \fig{departurecoefficient},
for the ground states of the two 
\markaschanged{ionisation} stages,
as well as for the levels we listed
in \tab{abundancelines}.
The plots reveal a generally smooth trend 
in the departure coefficients as a function of excitation energy.
In the line-forming regions, the low-excitation
\SiI~levels become overpopulated
while the high-excitation \SiI~levels 
and the ground state of \SiII~become underpopulated
relative to their Saha-Boltzmann equilibrium populations.

The picture presented by the departure
coefficients in \fig{departurecoefficient}~is characteristic
of photon suction \citep{2005ARA&amp;A..43..481A}:
photon losses in the \SiI~lines
drive a downward flow from the high-excitation levels to
the low-excitation levels.
Efficient coupling with the ground state of \SiII~(mainly 
mediated by charge transfer with neutral hydrogen; 
\fig{eqwtest})
transfers population
from the majority \SiII~species into the minority \SiI~species,
which further fuels the photon suction.
The net effect is that the \SiII~species
becomes slightly underpopulated in the upper atmosphere while
the \SiI~species as a whole becomes overpopulated
relative to their Saha 
\markaschanged{ionisation} equilibrium populations;
furthermore the lower levels of \SiI~become
overpopulated while the higher levels of \SiI~become underpopulated
relative to their Boltzmann excitation equilibrium populations.

We now consider the effects on the emergent equivalent widths
of the lines we listed in \tab{abundancelines}.
To a good approximation, the line opacities go as $\beta_{\text{lower}}$~and
the line source functions go as the ratio 
$\beta_{\text{upper}}/\beta_{\text{lower}}$ 
\citep[e.g.][]{2003rtsa.book.....R}.
As the lower levels of the intermediate-excitation \SiI~lines
are overpopulated relative to their Saha-Boltzmann equilibrium populations,
while their upper levels
are underpopulated relative to their Saha-Boltzmann equilibrium populations,
it follows that these lines
are stronger in non-LTE than in LTE if given the same silicon abundance,
by virtue of both an opacity effect and source function effect.
In contrast, the lower and upper levels of the
low-excitation \SiI~lines have similar departure coefficients
in the line-forming regions:
these lines are slightly 
stronger in non-LTE than in LTE if given the same silicon abundance,
but only by virtue of an opacity effect.

We illustrate this line-strengthening effect 
\markaschanged{in the vertical intensity}
across the surface of a single snapshot in \fig{map}.
The departures from LTE are most severe in the intergranular 
\markaschanged{downflows},
a characteristic feature of photon losses
\citep[e.g.][]{2016MNRAS.455.3735A}.
In contrast, in the \markaschanged{granular upflows} the 
ratio of the non-LTE to LTE equivalent widths are close to 
\markaschanged{unity.
Since the granular upflows have the larger
filling factor \citep[by roughly a factor of two;
e.g.][]{2013A&amp;A...557A..26M},
the overall non-LTE effect on the equivalent widths is small.}

\subsection{3D versus \mtd}
\label{results3d}

It is interesting to briefly compare the average results obtained 
from the full 3D model at great computational cost,
with those obtained from the average \mtd~model
at low computational cost. Before doing so 
we emphasize that all evidence suggests a full 3D analysis
is required to obtain \markaschanged{the highest accuracy},
as gauged by spectral line \markaschanged{strengths}, 
shapes and continuum fluctuations across the solar disk
\markaschanged{\citep{2009A&amp;A...507..417P,2013A&amp;A...554A.118P,
2011ApJ...736...69U}}.
However, it may be that non-LTE calculations using 
\mtd~model atmospheres present 
a reasonable compromise between computational cost 
and accuracy.

\markaschanged{Qualitatively,}
the contribution functions in 
\fig{contributionfunction}~\markaschanged{behave similarly} in the \mtd~model 
\markaschanged{as} in the 
3D model, for all of the lines considered in this study.
This indicates that \SiI~line-forming regions are similar
in the \mtd~model atmosphere as in \markaschanged{the 3D model atmosphere}.
The departure coefficients in \fig{departurecoefficient}~\markaschanged{also}
show the same qualitative behaviour
in the \mtd~model atmosphere as in the 3D model atmosphere,
\markaschanged{at least in the layers 
$\lgt\gtrsim-3$.}
\markaschanged{Quantitatively, we} 
found that the \mtd~LTE equivalent widths
are typically about $0.01\,\mathrm{dex}$~stronger
than the corresponding 3D LTE equivalent widths
for the intermediate-excitation \SiI~lines.
The \mtd~non-LTE equivalent widths
are also typically stronger 
than the corresponding 3D non-LTE equivalent widths,
and by a similar amount. 
This suggests a reasonably cheap and accurate approach, 
as compared to the detailed 3D non-LTE radiative transfer approach,
may be to apply \mtd~non-LTE versus \mtd~LTE abundance corrections
to 3D LTE results.
Before committing to this statement however,
we shall have to perform similar investigations for other species,
exhibiting different non-LTE effects, and for
other types of stars.

\subsection{Relative importance of radiative and collisional transitions} 
\label{resultssensitivity}
\begin{figure}
\begin{center}
\includegraphics[scale=0.3]{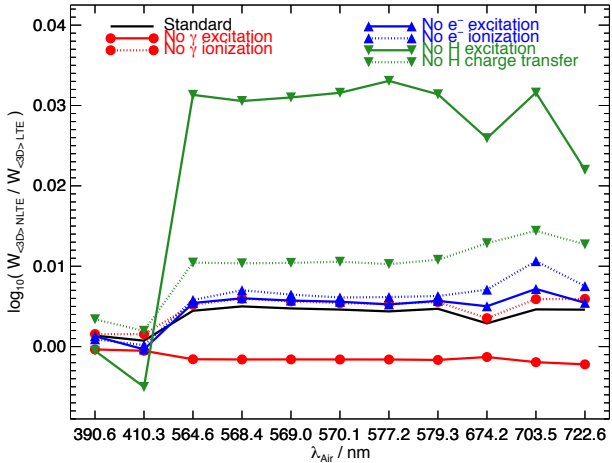}
\caption{Non-LTE to LTE equivalent width ratios
\markaschanged{in the vertical intensity},
in the \mtd~model solar atmosphere for
a fixed silicon abundance $\lgeps{Si}=7.51$,
when different radiative and collisional 
transitions are switched off.
A larger ratio indicates a strengthening of the line
in non-LTE
(the LTE line widths are identical for each
of the cases considered in this plot; 
hence this choice of normalization),
and thus that the abundance
corrections would be more negative.}
\label{eqwtest}
\end{center}
\end{figure}

We illustrate the relative importance of the different radiative and
collisional processes in \fig{eqwtest}, which shows
the non-LTE to LTE equivalent width ratios
for the $11\,\mathrm{lines}$~in the \mtd~model solar atmosphere for
a fixed silicon abundance $\lgeps{Si}=7.51$,
when different radiative and collisional 
transitions are switched off.

Photon losses in the \SiI~lines are what drive
the photon suction non-LTE effect we described in \sect{resultsnonlte}.
Switching off all of the 
radiative bound-bound transitions in the statistical equilibrium
calculations, impairs the population flow downwards 
from the higher levels of \SiI~into the lower levels of \SiI.
The level populations then become close to their Saha-Boltzmann equilibrium
values in the line-forming regions, and,
as we show in \fig{eqwtest}, the non-LTE effects
in the \SiI~lines become negligible.

Collisional excitation has important thermalizing effects
on the statistical equilibrium.
Both electron collisions and neutral hydrogen collisions
are efficient for transitions involving high-excitation \SiI~levels,
ensuring their relative populations are given by Boltzmann statistics.
For transitions involving low- and intermediate-excitation \SiI~levels,
electron collisions are more important in the optically-thick
atmosphere, whereas neutral hydrogen collisions are more important
higher up in the \SiI~line-forming regions,
where the hydrogen-to-free electron number ratio is larger.
The neutral hydrogen collisions provide a 
key opposition to the effects of photon suction
in the \SiI~lines of interest:
as we show in \fig{eqwtest},
neglecting these collisions would 
significantly increase the predicted non-LTE effects,
by about $0.03\,\mathrm{dex}$~in the solar case.
Neglecting electron collisions while retaining
neutral hydrogen collisions
has only a perturbative effect on the line strengths,
highlighting the dominance of the neutral hydrogen collisions.

Collisional \markaschanged{ionisation}
has a significant effect on the
statistical equilibrium in the line forming regions;
as with collisional excitation, neutral hydrogen collisions
have more effect on the final results than 
the electron collisions.
Charge transfer reactions with neutral hydrogen
efficiently link the intermediate-excitation 
\SiI~levels to the ground state of \SiII:
this reduces the overpopulation of these \SiI~levels
and reduces the underpopulation of the \SiII~level,
and brings the whole system closer to Saha-Boltzmann equilibrium.
As we show in \fig{eqwtest}, the non-LTE effects
are systematically larger when charge transfer reactions are neglected.
In contrast, \markaschanged{ionising} electron collisions
are more efficient for the high-excitation \SiI~levels.
Their effect on the statistical equilibrium is complicated.
They reduce the underpopulation of the
higher \SiI~levels, bringing them closer to
Saha-Boltzmann equilibrium,
at the cost of
directly increasing the underpopulation of the \SiII~level,
taking it further from  
Saha-Boltzmann equilibrium.
This indirectly increases
the overpopulation of the lower \SiI~levels
by providing fuel for the photon suction effects.
Consequently, while the non-LTE effects
in the high-excitation \SiI~lines
are larger when electron 
\markaschanged{ionisation} reactions are neglected,
they are smaller in the low-excitation \SiI~lines.

\markaschanged{Photoionisation} does not have a substantial impact on
the statistical equilibrium in the \SiI~line-forming regions,
in contrast to the findings of \citet{2008A&amp;A...486..303S}.
This can perhaps be attributed to our 
inclusion of charge transfer by neutral hydrogen, 
which acts to balance
to any potential 
\markaschanged{overionisation} or over-recombination effects.

\section{Solar photospheric silicon abundance}
\label{discussion}
\begin{table*}
\begin{center}
\caption{3D non-LTE solar photospheric silicon abundances
($\lgeps{3N}$)~inferred from the nine high-excitation optical 
\markaschanged{\SiI~lines}.
The abundance corrections $\Delta_{\mathrm{3N-3L}}$~are based on the 
theoretical equivalent widths that are consistent with the 
3D LTE abundances ($\lgeps{3L}$)
that were inferred by \citet{2015A&amp;A...573A..25S} 
in the
\markaschanged{solar disk-centre intensity spectrum}.
Also shown is the single
\SiII~line at $637.137\,\mathrm{nm}$~that was
analysed by \citet{2015A&amp;A...573A..25S}.
The \SiII~line was included in the weighted mean
used to obtain a final solar photospheric silicon abundance,
under the assumption that it forms in LTE conditions
(the derivation of abundance corrections for 
\SiII~lines is beyond the scope of this work).
\markaschanged{The equivalent widths are those
measured in the
solar disk-centre intensity spectrum
by \citet{2015A&amp;A...573A..25S}, based on the
average value obtained from the Kitt Peak atlas
\citep{1984SoPh...90..205N,brault1987spectral,1999SoPh..184..421N}~
and the Jungfraujoch atlas
\citep{1973apds.book.....D,1995ASPC...81...32D}.}}
\label{abundancecorrections}
\begin{tabular}{c c c c c c c}
\hline
Species &
$\lambda_{\text{Air}} / \mathrm{nm}$ &
Weight & 
$W / \mathrm{pm}$ &
$\lgeps{3L}$ &
$\Delta_{\mathrm{3N-3L}}$ &
$\lgeps{3N}$ \\
\hline
\hline
\SiI &
$564.561$ &
$1.0$ &
$3.50$ &
$7.507$ &
$-0.008$ &
$7.499$ \\
 
\SiI &
$568.448$ &
$2.0$ &
$6.37$ &
$7.464$ &
$-0.013$ &
$7.451$ \\
 
\SiI &
$569.043$ &
$3.0$ &
$5.26$ &
$7.510$ &
$-0.011$ &
$7.499$ \\
 
\SiI &
$570.111$ &
$3.0$ &
$3.95$ &
$7.491$ &
$-0.009$ &
$7.482$ \\
 
\SiI &
$577.215$ &
$2.0$ &
$5.60$ &
$7.563$ &
$-0.010$ &
$7.553$ \\
 
\SiI &
$579.307$ &
$1.0$ &
$4.58$ &
$7.601$ &
$-0.010$ &
$7.591$ \\
 
\SiI &
$674.164$ &
$1.0$ &
$1.63$ &
$7.614$ &
$-0.004$ &
$7.610$ \\
 
\SiI &
$703.490$ &
$1.0$ &
$7.40$ &
$7.552$ &
$-0.011$ &
$7.541$ \\
 
\SiI &
$722.621$ &
$1.0$ &
$3.87$ &
$7.508$ &
$-0.009$ &
$7.499$ \\
 
\hline
\SiII &
$637.137$ &
$1.0$ &
$3.66$ &
$7.539$ &
$\left(-0.000\right)$ &
$\left(7.539\right)$ \\
\hline
\hline
\end{tabular}
\end{center}
\end{table*}

\begin{figure}
\begin{center}
\includegraphics[scale=0.3]{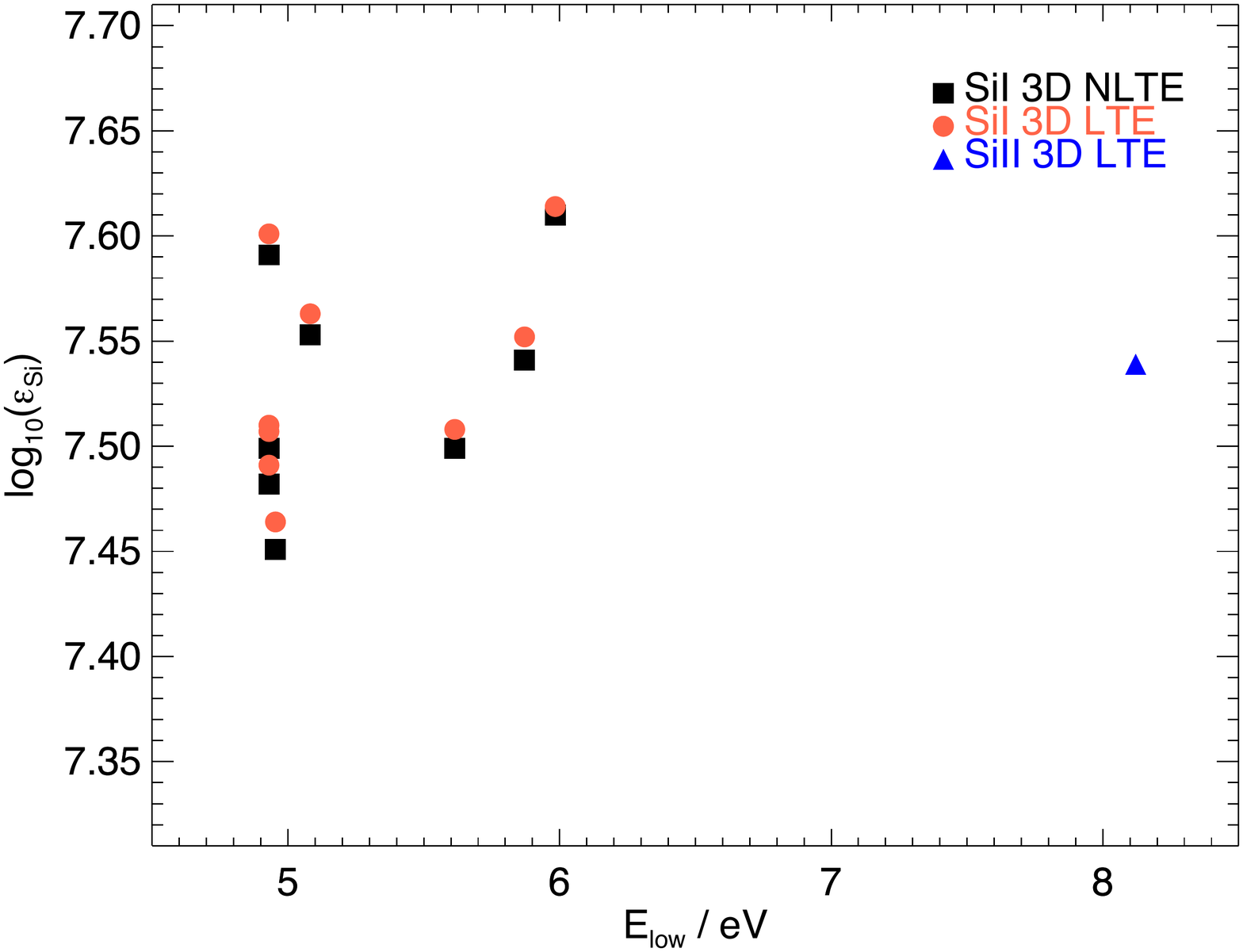}
\includegraphics[scale=0.3]{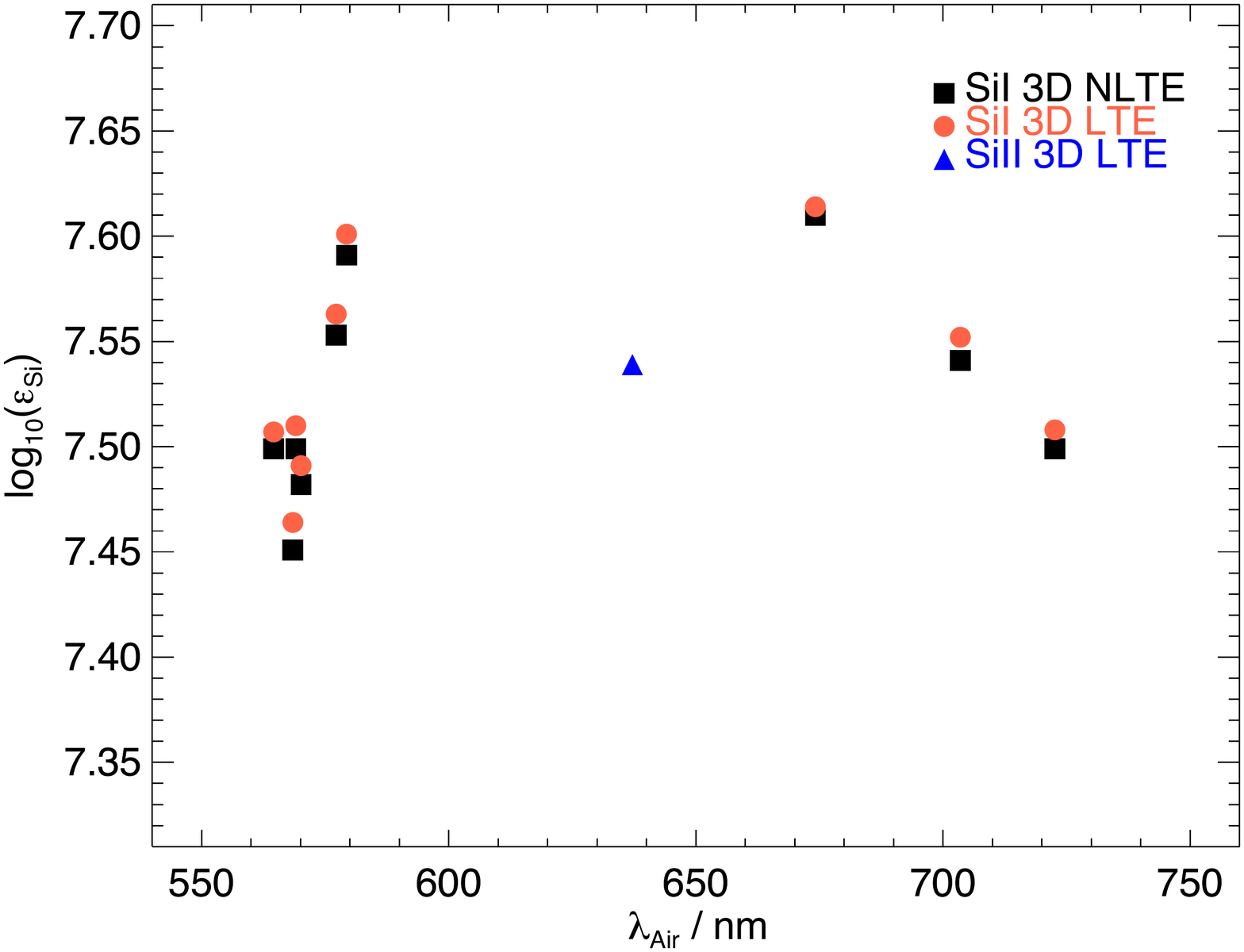}
\includegraphics[scale=0.3]{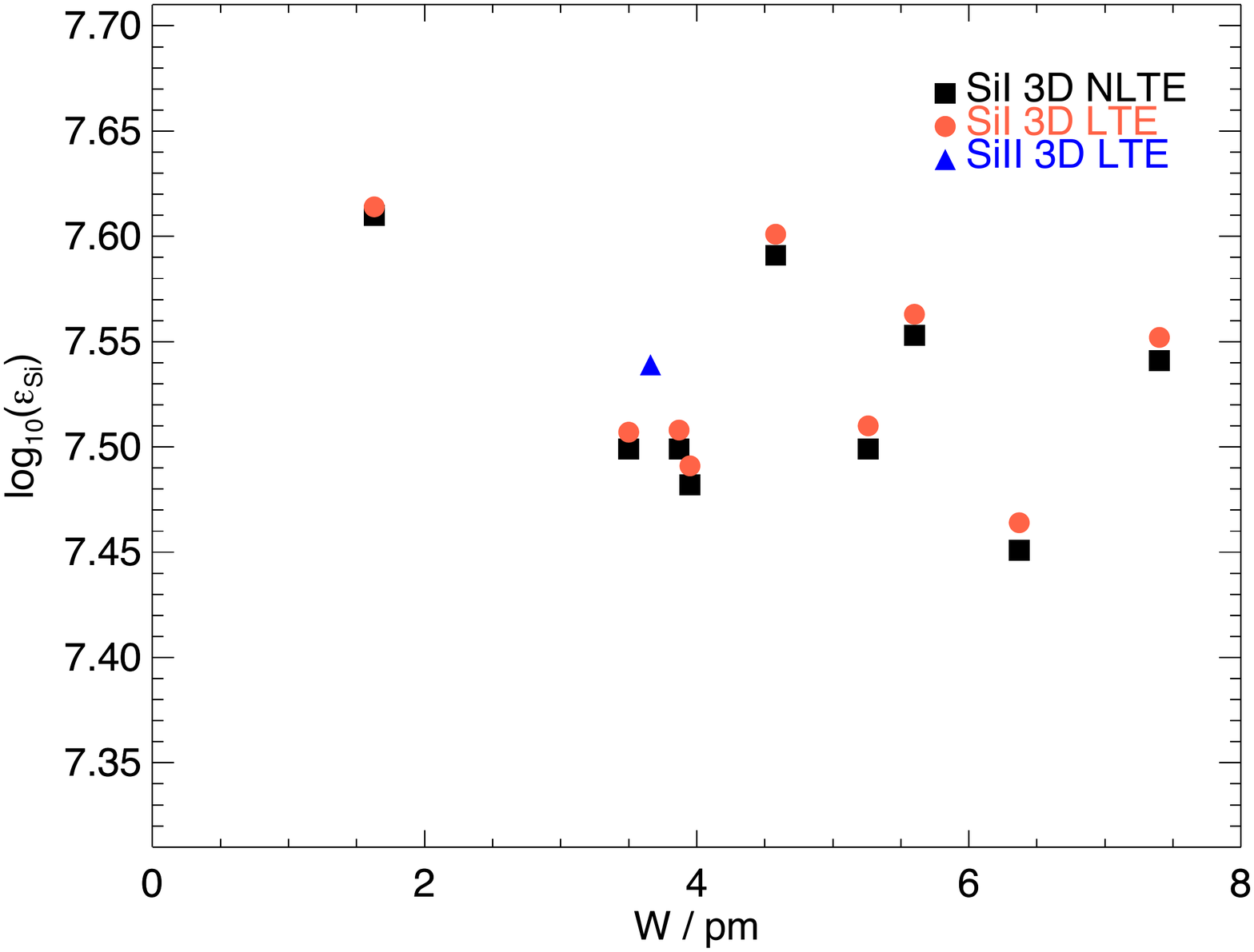}
\caption{\markaschanged{Solar 
silicon abundances inferred from the individual lines,
as functions of: the excitation potential of the lower level of the lines;
the wavelength of the lines; and the equivalent widths of the lines
measured in the solar disk-centre intensity spectrum
by \citet{2015A&amp;A...573A..25S}.}}
\label{trends}
\end{center}
\end{figure}

We derived 3D non-LTE versus 3D LTE 
abundance corrections for each of the
nine \SiI~lines analysed by \citet{2015A&amp;A...573A..25S}~in 
the solar disk-centre intensity spectrum.
For a given line, this was done by finding the 
theoretical equivalent width consistent with the 
3D LTE abundance ($\lgeps{3L}$)~inferred 
by \citet{2015A&amp;A...573A..25S},
and subsequently finding \markaschanged{the 3D non-LTE abundance}
($\lgeps{3N}$) that
corresponds to this equivalent width.
The abundance correction is then
$\Delta_{\mathrm{3N-3L}}=\lgeps{3N}-\lgeps{3L}$.
This somewhat roundabout way is necessary since 
here we only average over six snapshots,
while the 3D LTE study of \citet{2015A&amp;A...573A..25S}~has
a very fine sampling based on ninety snapshots
across the same temporal range.
We list the abundance corrections 
and inferred 3D non-LTE abundances
for the individual lines in \tab{abundancecorrections}
\markaschanged{and display them graphically as functions of the
line parameters in \fig{trends}.}
Our 3D non-LTE abundance corrections
for the \markaschanged{disk-centre} 
intensity spectrum are very similar to the corresponding 
1D non-LTE corrections for the solar flux spectrum of 
\citet{2008A&amp;A...486..303S} that were
adopted by \citet{2015A&amp;A...573A..25S}.

To obtain \markaschanged{an estimate}
for the solar photospheric silicon abundance,
a weighted mean was computed using 
our nine 3D non-LTE abundances from \SiI~lines,
as well as the single \SiII~line analysed 
by \citet{2015A&amp;A...573A..25S},
\citep[assuming that the 3D non-LTE abundance corrections for
this latter line is zero
as previous work using 1D radiative transfer has suggested;
e.g.][]{2008A&amp;A...486..303S,2013ApJ...764..115B}.
Using the weights of \citet{2015A&amp;A...573A..25S},
we obtain $\lgeps{Si\odot}=7.51\pm0.03\,\mathrm{dex}$,
adopting the same uncertainty as
\citet{2015A&amp;A...573A..25S}~that
includes both statistical and systematic errors.

Our inferred solar photospheric silicon abundance is
the same as advocated by 
\citet{2000A&amp;A...359..755A},
\citet{2005ASPC..336...25A},
\citet{2009ARA&amp;A..47..481A},
and \citet{2015A&amp;A...573A..25S}.
This is reassuring; as discussed in 
\citet{2009ARA&amp;A..47..481A},
the canonical photospheric abundances 
are in excellent overall agreement 
with the corresponding meteoritic abundances
\citep{2009LanB...4B...44L}
when adopting this silicon abundance.


\section{Conclusion}
\label{conclusion}

We have presented 3D non-LTE \SiI~line formation calculations
using the 3D hydrodynamic \stagger~model solar atmosphere
of \citet{2009ARA&amp;A..47..481A}~and
using a realistic model atom that includes recent 
quantum-mechanical neutral hydrogen collision data
from \citet{2014A&amp;A...572A.103B}.
Our main findings are:
\begin{itemize}
\item{The non-LTE effect on the level populations
is that of photon suction: photon losses in the \SiI~lines
drive a population flow downwards, such that
the lower levels are overpopulated
and the higher levels are underpopulated,
relative to their Saha-Boltzmann equilibrium populations.}
\item{The non-LTE effects on the emergent equivalent widths
are largest in the \markaschanged{intergranular downflows}, 
and nearly negligible in the 
\markaschanged{granular upflows}. 
The larger filling factor of the
\markaschanged{granular upflows make} 
the overall 3D non-LTE versus 3D LTE 
abundance corrections are close to zero.}
\item{We confirm the result of \citet{2016AstL...42..366M},
that collisions with neutral hydrogen
have a strong thermalizing effect on the
statistical equilibrium.
Excitation reactions provide a particularly important
opposition to the photon suction effect.}
\item{Applying our derived 3D non-LTE versus 3D LTE  
abundance corrections line-by-line to the 3D LTE solar photospheric
silicon abundances derived by \citet{2015A&amp;A...573A..25S},
we infer a 3D non-LTE 
solar photospheric silicon abundance of $7.51\,\mathrm{dex}$.
This is the same as the current canonical value
of \citet{2009ARA&amp;A..47..481A}~and
\citet{2015A&amp;A...573A..25S},
which had adopted 1D non-LTE abundance corrections.}
\end{itemize}

We anticipate that 
future work on the solar photospheric chemical composition
will increasingly utilize 3D non-LTE~radiative transfer 
techniques such as those discussed in this paper;
this is a necessary development for reducing
the systematic modelling uncertainties to a level
comparable to or better than the 
\markaschanged{uncertainties} associated with
the observations and, in particular, the 
oscillator strengths of the diagnostic lines.
Beyond the Sun, future theoretical work on neutral silicon may focus
on the \SiI~$390.6\,\mathrm{nm}$~and $410.3\,\mathrm{nm}$~lines
in the metal-poor regime ($\feh\lesssim-2$), where
strong UV \markaschanged{overionisation} effects are 
likely to drive positive
non-LTE versus LTE abundance corrections of up to around
$0.2\,\mathrm{dex}$.

\section*{Acknowledgements}
\label{acknowledgements}
The authors thank the anonymous referee for their constructive feedback
on the original manuscript.
The authors are supported by the Australian
Research Council (ARC) grant FL110100012.
This research was undertaken with the 
assistance of resources from the 
National Computational Infrastructure (NCI),
which is supported by the Australian Government.

\bibliographystyle{mnras}
\bibliography{/Users/ama51/Documents/work/papers/allpapers/bibl.bib}

\label{lastpage}
\end{document}